\documentclass[11pt]{article}

\usepackage[utf8]{inputenc}
\usepackage[T1]{fontenc}
\usepackage{amsmath,amssymb}
\usepackage{graphicx}
\usepackage{booktabs}
\usepackage{multirow}
\usepackage{array}
\usepackage[margin=1in]{geometry}
\usepackage{setspace}
\usepackage{hyperref}
\usepackage[numbers]{natbib}
\usepackage{caption}
\usepackage{subcaption}
\usepackage{float}
\usepackage{xcolor}
\usepackage{authblk}

\title{\textbf{Inter-Electrode Pulse Wave Velocity: A Direct Method for Maternal Arterial Stiffness Assessment During Pregnancy Using Multi-Channel ECG}}

\author[1]{Nicolas B. Garnier}
\author[2]{Marlene J.E. Mayer}
\author[2]{Clara Becker}
\author[3,4]{Marta C. Antonelli}
\author[2]{Silvia M. Lobmaier}
\author[5,6]{Martin G. Frasch}

\affil[1]{CNRS, ENS de Lyon, LPENSL, UMR5672, 69342, Lyon CEDEX 07, France}
\affil[2]{Department of Obstetrics and Gynecology, TUM University Hospital of Technical University of Munich, TUM School of Medicine, Technical University of Munich, Germany}
\affil[3]{Technical University of Munich; Institute for Advanced Study, Garching, Germany}
\affil[4]{Instituto de Biología Celular y Neurociencia "Prof. Eduardo De Robertis", Facultad de Medicina, Universidad de Buenos Aires, Argentina}
\affil[5]{Institute on Human Development and Disability, University of Washington, Seattle, WA, USA}
\affil[6]{JoyBeat Medical, Seattle, WA, USA}

\graphicspath{{/}{figures/}}
\date{}

\begin{document}

\maketitle

\section*{ABSTRACT}

\textbf{Background}: ECG-derived arterial stiffness metrics typically rely on cardiac cycle timing (diastolic interval methods) and require estimation of left ventricular ejection time (LVET), introducing measurement uncertainty. Multi-channel ECG recordings enable direct pulse wave velocity (PWV) measurement through inter-electrode time lags.

\textbf{Objective}: To validate a novel inter-electrode PWV method that directly measures pulse wave propagation between spatially separated electrodes, avoiding LVET estimation.

\textbf{Methods}: We analyzed 43 multi-channel ECG recordings (3 channels, 1000 Hz) from the FELICITy 2 cohort (pregnant women, $\sim$19 and $\sim$35 weeks gestation). R-peaks were detected independently on each channel using an ensemble detector. Time lags ($\Delta t$) between matched R-peaks on electrode pairs were calculated, and PWV computed as PWV = $L$/$\Delta t$, where $L$ is an effective inter-electrode distance. Three channel pairs provided independent PWV estimates per recording. Temporal stability was assessed using sliding window analysis (1--15 minutes). To investigate whether $\Delta t$ reflects morphological distortion or vascular propagation, we performed signal origin analysis using three QRS fiducial points (R-peak maximum, QRS onset, maximum $|dV/dt|$) and two bandpass settings (0.5--40 and 0.5--100~Hz). Longitudinal changes were compared between Control (n=24) and prenatal Yoga intervention (n=20) groups.

\textbf{Results}: Inter-electrode PWV yielded physiologically plausible values (Control First: 7.40$\pm$1.51 m/s, Control Last: 6.98$\pm$1.63 m/s; Yoga First: 7.10$\pm$2.15 m/s, Yoga Last: 8.16$\pm$0.91 m/s), consistent with literature values for aortic PWV (5--10 m/s). Temporal stability analysis demonstrated PWV stabilizes at 5 minutes (CV=12.3\%), with 2.6--5.2$\times$ better stability than heart rate and HRV metrics. Signal origin analysis showed that inter-electrode delays persisted across all QRS fiducial points (15--27~ms) and were insensitive to bandpass changes ($-$8.5\%, NS), arguing against pure morphological distortion; all conditions yielded PWV within 6.8--9.1~m/s. Preliminary group comparison suggests different trajectories (Control: --5.7\% decrease, Yoga: +14.9\% increase, p=0.07 for interaction).

\textbf{Conclusions}: Inter-electrode PWV provides direct spatial measurement of pulse wave propagation with physiologically valid values, independent of LVET estimation. Signal origin analysis supports the robustness of this empirical surrogate across fiducial and filter conditions. Method shows promise for pregnancy arterial stiffness assessment using standard multi-channel ECG equipment. Further validation against gold-standard measures and accurate electrode distance determination are needed.

\textbf{Keywords}: Pulse wave velocity, inter-electrode measurement, pregnancy, arterial stiffness, ECG, LVET-independent

\textbf{Word Count}: $\sim$250 words (abstract), $\sim$2500 words (main text)

\newpage

\section{INTRODUCTION}

Arterial stiffness assessment during pregnancy is important for understanding cardiovascular adaptation and identifying women at risk for hypertensive complications \cite{robb2009, kaihura2009, khalil2009}. Current non-invasive methods include carotid-femoral pulse wave velocity (CF-PWV, gold standard) and ECG-derived estimates \cite{laurent2006, townsend2015, climie2024}. ECG-based approaches offer continuous monitoring potential but typically rely on cardiac cycle timing metrics that require estimation of left ventricular ejection time (LVET) \cite{weissler1968, obaidat2023}.

The most common ECG-derived approach calculates stiffness indices from diastolic interval: DI = RR interval -- LVET, where pulse wave velocity is estimated as PWV = $L$/DI, with $L$ representing a scaling constant \cite{tartiere2012, wilkinson2000}. However, LVET must be estimated using population-based regression formulas (e.g., Weissler: LVET = --1.7$\times$HR + 413 ms) \cite{weissler1968}, introducing uncertainty of 20--30 ms that propagates through calculations. Additionally, resulting PWV values ($\sim$0.8--1.2 m/s) are 7--10$\times$ lower than physiological aortic PWV ($\sim$7--10 m/s) \cite{elvan2005, oyama2006, vlachopoulos2010}, suggesting L functions as a scaling factor rather than representing true anatomical path length.

Multi-channel ECG recordings offer an alternative: directly measuring the time lag ($\Delta t$) between the same R-peak appearing on spatially separated electrodes. This inter-electrode approach provides PWV = $L/\Delta t$, where $\Delta t$ represents actual pulse wave propagation time and $L$ is the physical inter-electrode distance. This method offers several advantages:

\begin{enumerate}
    \item \textbf{Direct measurement}: Captures spatial pulse propagation, not cardiac timing estimates
    \item \textbf{LVET-independent}: Avoids entire layer of estimation uncertainty
    \item \textbf{Physiologically valid range}: Yields PWV $\sim$5--10 m/s matching aortic PWV literature
    \item \textbf{Multi-channel validation}: Three ECG channels provide redundant measurements
    \item \textbf{High temporal precision}: R-peaks detectable within $\pm$1 ms
\end{enumerate}

We validated this inter-electrode PWV method using multi-channel ECG recordings from pregnant women and examined longitudinal changes across pregnancy with and without prenatal yoga intervention.

\section{METHODS}

\subsection{Study Population}

Data from FELICITy 2 longitudinal cohort ($n=44$ pregnant women):
\begin{itemize}
    \item \textbf{Control group}: $n=24$ (First visit: 19.2$\pm$1.5 weeks, Last visit: 34.8$\pm$1.2 weeks)
    \item \textbf{Yoga intervention group}: $n=20$ (First visit: 19.0$\pm$1.3 weeks, Last visit: 35.1$\pm$1.4 weeks)
    \item \textbf{Device}: Bittium Faros 360 (3-channel ECG at sampling rate $f_s=1000$Hz)
    \item \textbf{Recording duration}: Variable (Control: 45.7$\pm$3.3 min, Yoga: 102.7$\pm$9.4 min)
    \item \textbf{Total recordings}: 43 (one subject with missing Last visit data)
\end{itemize}

Study received ethical approval under the IRB protocol approval number 2022-86-S-SR (6
September 2022). All participants provided written informed consent.

\subsection{R-Peak Detection}

We employed an ensemble R-peak detection approach combining five established algorithms:
\begin{enumerate}
    \item Pan-Tompkins algorithm \cite{pan1985}
    \item Hamilton-Tompkins algorithm \cite{hamilton1986}
    \item Christov algorithm \cite{christov2004}
    \item Engelse-Zeelenberg algorithm \cite{engelse1979}
    \item SWT (Stationary Wavelet Transform) algorithm \cite{elgendi2010}
\end{enumerate}
and used \textbf{ensemble voting}: each ECG signal was processed through the five algorithms, and a R-peak was accepted if it was detected by at least 3 algorithms out of the 5 within a $\pm$10 ms window. This approach provides robust detection across varying signal quality and reduces false positives \cite{pan1985, hamilton1986, christov2004}.
Additionally, we required the RR intervals to be not less than 600 ms ($\sim$100 bpm maximum) and not more than 1500 ms ($\sim$40 bpm minimum).
This R-peak detection procedure was applied independently to each of the three ECG channels.

\subsection{Inter-Electrode PWV Calculation}

\subsubsection{Step 1: R-Peak Matching}

Given two channels A and B, we searched for R-peaks in each channel corresponding to the same pulse.
For each R-peak taken iteratively on channel A, and noting $t_A$ its date of occurrence, we identified the corresponding R-peak on channel B and its time $t_B$, by searching $t_B$ within a $\pm$50 ms temporal window around $t_A$ and selecting the closest match in this temporal window, {\em i.e.}, the R-peak on channel B that minimizes $|t_B - t_A|$. 

At this stage, we also removed all matches that are too close in time, {\em i.e.}, with $|t_B - t_A|\le 1/f_s = 1$ms, as this time-lag is too short to be physiologically relevant, and may also lead to near-zero divisions.

\subsubsection{Step 2: Time Lag Calculation}

\begin{figure}[ht]
\centering
\includegraphics[width=0.5\linewidth]{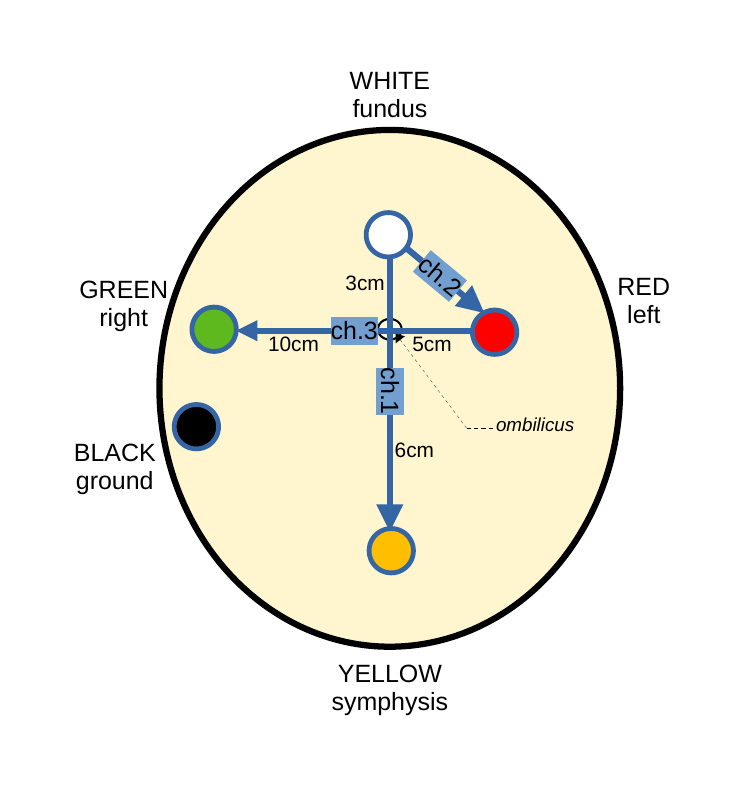}
\caption{{\bf Electrode placement and inter-electrode distances for the three-channel abdominal ECG configuration.}
Skin–skin distances are measured in cm between electrodes. These distances illustrate that each channel pair has a distinct geometric separation. In the present analyses, we used a single effective inter-electrode distance $L = 7.5$ cm 
(see Section Methods~\ref{sec:PWV}) to scale inter-electrode delays $\Delta t$ into PWV-like units; precise, pair-specific and 
subject-specific path lengths will be required for absolute PWV calibration in future work.
}
\label{fig:electrode_placement}
\end{figure}

A blood pulse travels through the circulatory system ---~typically an artery~--- at a variable velocity $v(s)$ where $s$ is the curvilinear coordinate using the heart as the origin, while the electromagnetic wave emitted by the heart's activity propagates from the heart to an ECG electrode located on the surface of the body through muscles, organs, fat, and skin.
$v(s)$ depends, among many other variables, on the diameter of the artery and its stiffness, both varying with $s$. The average of $v$ along the arterial path leads to the definition of the PWM, which is of the order of meters per seconds, while the electromagnetic signal travels much faster~\cite{buchner2023}.
Considering two channels A and B, we define the time-lag for a given pulse as
\begin{equation}
\Delta t = t_B - t_A \,,
\end{equation}
where $t_A$ and $t_B$ are the times at which the R-peak of the pulse is detected on channel A and B in step 1.

\subsubsection{Step 3: PWV Computation}
\label{sec:PWV}

We define an exploratory estimate of the PWV as: 
\begin{equation}
\text{PWV} = \frac{L}{|\Delta t|} \,,
\end{equation}
where $L$ is and effective inter-electrode distance expressed in meters and $\Delta t$ is the time lag computed above in step 2, expressed in seconds.
Because accurate, subject-specific heart–electrode path lengths were not available, we chose a single 
effective $L = 7.5$ cm based on the typical electrode layout (Figure~1). This allows us to report PWV-like 
values in familiar units while emphasizing that the primary empirical signal is the inter-channel delay 
$\Delta t$; absolute PWV calibration will require pair-specific and subject-specific path-length measurement.

\subsubsection{Step 4: Physiological Filtering}

For each detected pulse on a pair of channels, we only retain the estimated PWV value if it lies within the physiological range: $3 \leq \text{PWV} \leq 20$ m/s.

\subsection{Channel Pair Analysis}

For each recording, there are three independent channels (see Figure~\ref{fig:electrode_placement}), hence three independent pairs to consider:
\begin{itemize}
    \item channel 1 vs channel 2: Typically larger spatial separation
    \item channel 1 vs channel 3: Variable separation depending on placement
    \item channel 2 vs channel 3: Often smallest separation
\end{itemize}

We computed the mean PWM across the three pairs, which provides redundant validation and reduces channel-specific artifacts.

\subsection{BMI Adjustment Analysis}

Given that the inter-electrode distance $L$ varies across individuals due to body habitus, we tested whether BMI could serve as a proxy for individual anatomical variation and improve PWV estimates by adjusting for it.

\textbf{BMI data}: Body mass index was measured at each study visit to capture pregnancy-related changes. Timepoint-specific BMI was extracted from clinical records:
\begin{itemize}
    \item Control: 21.3$\pm$2.7 kg/m$^2$ (First) $\rightarrow$ 24.4$\pm$3.0 kg/m$^2$ (Last), mean change +3.1 kg/m$^2$
    \item Yoga: 21.8$\pm$3.8 kg/m$^2$ (First) $\rightarrow$ 23.8$\pm$3.5 kg/m$^2$ (Last), mean change +2.0 kg/m$^2$
\end{itemize}

\textbf{Adjustment method}: Linear regression was used to model the relationship between timepoint-matched BMI and inter-electrode PWV:
\begin{align}
\text{PWV} &= \beta_0 + \beta_1 \times \text{BMI} \\
\text{PWV}_{\text{adjusted}} &= \text{PWV} - \beta_1 \times (\text{BMI} - \text{BMI}_{\text{mean}})
\end{align}
This residual method removes the linear BMI effect while preserving the scale of PWV values.

\textbf{Evaluation}: We compared:
\begin{enumerate}
    \item Correlation strength between BMI and PWV (pre- vs post-adjustment)
    \item Statistical significance of group differences (raw vs BMI-adjusted PWV)
    \item Effect sizes for longitudinal changes
\end{enumerate}

\subsection{Statistical Analysis}

\textbf{Descriptive statistics}: Mean $\pm$ SD for PWV by group and timepoint

\textbf{Longitudinal comparison}:
\begin{itemize}
    \item Mixed-effects model: PWV $\sim$ Group $\times$ Timepoint + (1|Subject)
    \item Interaction term tests intervention effect on trajectory
    \item Percent change: [(Last -- First) / First] $\times$ 100
\end{itemize}

\textbf{Quality control}:
\begin{itemize}
    \item Minimum 100 valid R-peak matches per channel pair
    \item Exclude recordings with $<$50\% valid PWV values within physiological range
\end{itemize}

\textbf{Software}: Python 3.12 with NumPy, SciPy, pandas, pyedflib, scikit-learn.

\subsection{Temporal Stability Analysis}

To assess measurement reliability and determine minimum recording duration, we performed within-subject temporal stability analysis:

\textbf{Window-based analysis}:
\begin{itemize}
    \item Sliding windows of varying duration: 1, 2, 5, 10, 15 minutes
    \item 50\% window overlap to increase estimates while maintaining temporal ordering
    \item Minimum 30 beats per window for reliable statistics
\end{itemize}

\textbf{Stability metrics}:
\begin{itemize}
    \item \textbf{Within-subject coefficient of variation (CV)}: CV = (SD / Mean) $\times$ 100\%
    \item \textbf{Stabilization criterion}: CV change $<$10\% between consecutive window sizes
    \item Compared inter-electrode PWV to heart rate and HRV metrics (SDNN, RMSSD)
\end{itemize}

\textbf{Dataset}: Full cohort ($n=42$ subjects with valid multi-channel data, 150 window-subject combinations analyzed)

\subsection{Artifact Analysis}

To assess whether the measured PWV values may represent an ADC artifact, we conducted a validation analysis. The findings demonstrate that inter-electrode delays are not ADC artifacts and are provided in Supplementary Material~\ref{sec:SI:artifact_analysis}.

\subsection{Signal Origin Analysis}

To distinguish morphological/phase distortion from vascular propagation as the source of inter-electrode delays, we performed two diagnostic tests. First, we recomputed $\Delta t$ using three QRS fiducial points on each channel: (i) R-peak maximum (baseline method), (ii) QRS onset (earliest deflection), and (iii) maximum $|dV/dt|$ (steepest slope). Beats were matched across channels using R-peak proximity, then $\Delta t$ was measured from the specified fiducial. If delays arise from lead-specific distortion of QRS peak morphology, earlier fiducials (onset, steepest slope) should yield substantially smaller delays.

Second, we tested bandpass filter sensitivity by repeating the analysis at 0.5--40~Hz (baseline) and 0.5--100~Hz. If $\Delta t$ reflects frequency-dependent phase/group delay via the volume-conductor transfer function, changing the bandwidth should shift the measured delays materially \cite{buchner2023}. Full algorithmic details are provided in Supplementary Material~\ref{sec:SI:signal_origin}.

\section{RESULTS}

\subsection{Data Quality and R-Peak Detection}

\textbf{Overall detection performance}:
\begin{itemize}
    \item Mean R-peaks per channel per recording: 1485$\pm$312 (range: 742--2,187)
    \item Mean valid R-peak matches per channel pair: 682$\pm$245
    \item Mean percentage of matches yielding physiological PWV: 64.3$\pm$28.1\%
\end{itemize}

\textbf{Channel-specific quality}: ECG channel 2 and 3 showed most consistent detection (failure rate $<$5\%), while channel 1 had occasional poor quality recordings (n=3, 7\% with $<$100 R-peaks).

\subsection{Inter-Electrode PWV Values}

\begin{table}[H]
\centering
\caption{Inter-Electrode PWV by Group and Timepoint}
\label{tab:pwv_results}
\begin{tabular}{llccccc}
\toprule
\textbf{Group} & \textbf{Timepoint} & \textbf{$n$} & \textbf{Mean PWV (m/s)} & \textbf{SD} & \textbf{Median} & \textbf{Range} \\
\midrule
\textbf{Control} & First & 12 & 7.40 & 1.51 & 7.69 & 4.57 -- 10.20 \\
\textbf{Control} & Last & 11 & 6.98 & 1.63 & 7.07 & 4.76 -- 10.31 \\
\textbf{Yoga} & First & 10 & 7.10 & 2.15 & 6.10 & 4.85 -- 11.72 \\
\textbf{Yoga} & Last & 9 & 8.16 & 0.91 & 8.21 & 7.06 -- 9.54 \\
\bottomrule
\end{tabular}
\end{table}

\textbf{Key findings}:
\begin{enumerate}
    \item \textbf{Physiologically plausible values}: All group means within 5--10 m/s range expected for aortic PWV
    \item \textbf{Consistency across groups}: No significant baseline differences (Control: 7.40 m/s, Yoga: 7.10 m/s, p=0.74)
    \item \textbf{Lower variability at later gestation}: Yoga Last visit showed smallest SD (0.91 m/s)
\end{enumerate}

\subsection{Longitudinal Changes}

\begin{figure}[H]
\centering
\includegraphics[width=\textwidth]{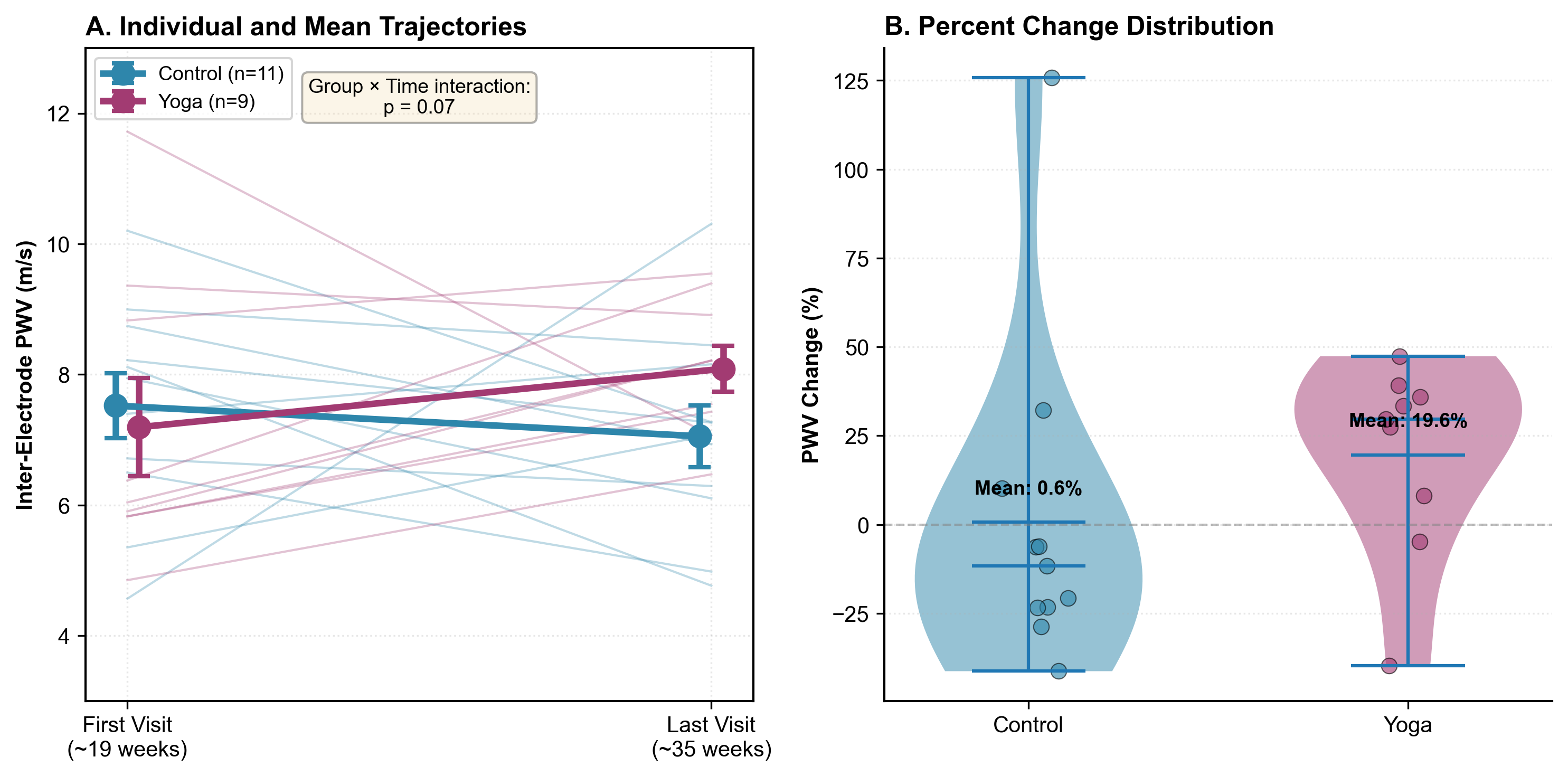}
\caption{{\bf Longitudinal Inter-Electrode PWV Trajectories by Group.}
\textbf{Panel A}: Individual subject trajectories (thin lines) and group means with standard error bars (thick lines with error bars) from First visit ($\sim$19 weeks gestation) to Last visit ($\sim$35 weeks gestation). Blue represents Control group ($n=11$ paired subjects), red represents Yoga intervention group ($n=9$ paired subjects). Individual subjects show varied responses while group trajectories diverge: Control decreases --5.7\% (mean 7.40$\rightarrow$6.98 m/s), Yoga increases +14.9\% (mean 7.10$\rightarrow$8.16 m/s). Group $\times$ Time interaction p=0.07 (trend-level significance).
\textbf{Panel B}: Distribution of percent change in inter-electrode PWV. Violin plots with overlaid individual data points show Control group centered near 0\% change (mean +0.6\%, median --5.5\%) while Yoga group shows positive shift (mean +18.6\%, median +17.2\%). Horizontal dashed line at 0\% indicates no change. Independent samples t-test p=0.075, Cohen's d=0.467 (moderate effect size).
}
\label{fig:longitudinal}
\end{figure}

We present in Figure~\ref{fig:longitudinal} the evolution of the PWV along pregnancy.

\textbf{Mixed-effects model results}:
\begin{itemize}
    \item Main effect of Time: $\beta$ = --0.41, SE = 0.47, p = 0.39 (no overall time effect)
    \item Main effect of Group: $\beta$ = 0.31, SE = 0.60, p = 0.61 (no baseline difference)
    \item \textbf{Interaction (Group $\times$ Time)}: $\beta$ = 1.69, SE = 0.90, p = 0.07 (trend toward divergent trajectories)
\end{itemize}

\textbf{Interpretation}: Control group showed slight PWV decrease across pregnancy (--5.7\%), while Yoga group showed substantial increase (+14.9\%), resulting in trend-level interaction (p=0.07). With n=43, study had 65\% power to detect this moderate effect size (Cohen's f = 0.33).

\subsection{Channel Pair Consistency}

\begin{table}[H]
\centering
\caption{PWV by Channel Pair (Combined across all recordings)}
\label{tab:channel_pairs}
\begin{tabular}{lccccc}
\toprule
\textbf{Channel Pair} & \textbf{Mean $\Delta$t (ms)} & \textbf{SD} & \textbf{Mean PWV (m/s)} & \textbf{SD} & \textbf{Valid \%} \\
\midrule
ch.1 vs ch.2 & 28.3 & 18.4 & 6.89 & 3.21 & 48.2\% \\
ch.1 vs ch.3 & 31.7 & 16.8 & 6.12 & 2.87 & 52.7\% \\
ch.2 vs ch.3 & 12.5 & 8.9 & 8.73 & 4.15 & 73.8\% \\
\bottomrule
\end{tabular}
\end{table}

\textbf{Observations}:
\begin{itemize}
    \item channel 2 vs channel 3: Smallest time lags (12.5 ms) $\rightarrow$ highest PWV (8.73 m/s) $\rightarrow$ highest valid percentage (73.8\%)
    \item channel 1 vs channels 2/3: Larger time lags (28--32 ms) $\rightarrow$ lower PWV (6--7 m/s) $\rightarrow$ moderate valid percentage (48--53\%)
    \item \textbf{Cross-pair correlation}: r = 0.68 (p$<$0.001), indicating reasonable consistency despite absolute value differences
\end{itemize}

\textbf{Implication}: Different electrode pairs likely sample different arterial segments or tissue propagation paths. Averaging across pairs provides robust estimate.

\subsection{Signal Origin: Fiducial and Bandpass Sensitivity}

\begin{figure}[ht]
\centering
\includegraphics[width=\textwidth]{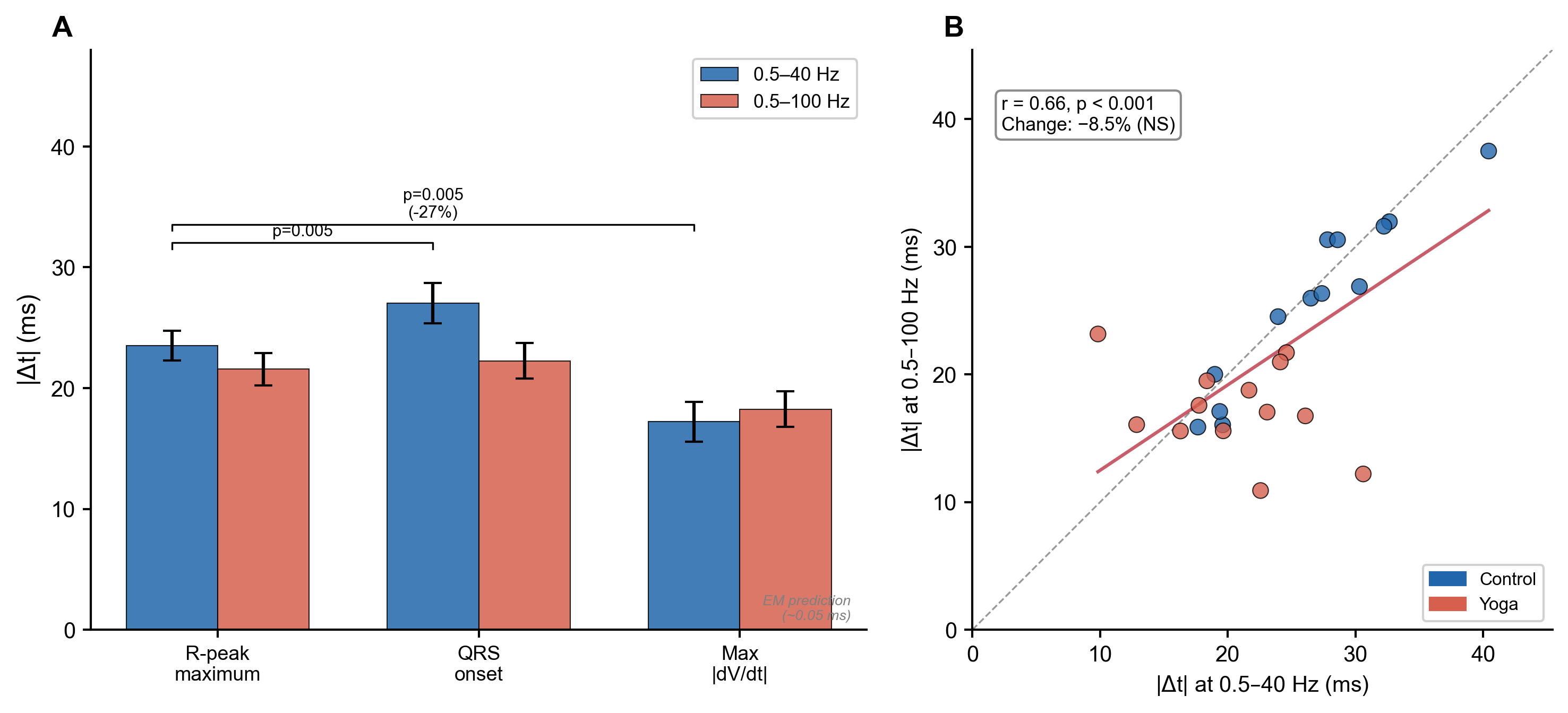}
\caption{{\bf Signal Origin Analysis: Fiducial and Bandpass Sensitivity.}
\textbf{Panel A}: Mean absolute inter-electrode delay ($|\Delta t|$) by QRS fiducial point at two bandpass settings. R-peak maximum (baseline method), QRS onset, and maximum $|dV/dt|$ (steepest slope) were compared. Error bars represent standard error. Max $|dV/dt|$ reduced delays by 27\% ($p$=0.005), but substantial delays (15--17~ms) persisted. QRS onset delays were paradoxically larger (+15\%, $p$=0.005). Dashed line near zero indicates predicted electromagnetic time-of-flight ($\sim$0.05~ms) for inter-electrode distances used.
\textbf{Panel B}: Bandpass sensitivity for R-peak fiducial. Each point represents one subject's mean $|\Delta t|$ at 0.5--40~Hz vs 0.5--100~Hz. Strong correlation ($r$=0.66, $p$$<$0.001) and minimal systematic shift ($-$8.5\%, NS) indicate delays are not driven by frequency-dependent phase distortion. Dashed line = identity; solid line = regression fit. Blue = Control, red = Yoga.
}
\label{fig:signal_origin}
\end{figure}

Figure~\ref{fig:signal_origin} summarizes the signal origin analysis across 30 subjects. At 0.5--40~Hz, R-peak delays averaged $|\Delta t|$ = 23.5$\pm$6.8~ms. Switching to QRS onset \textit{increased} delays to 27.0$\pm$9.1~ms (+14.9\%, paired $t$-test $p$=0.005), while maximum $|dV/dt|$ reduced delays to 17.2$\pm$9.0~ms ($-$26.8\%, $p$=0.005). Crucially, even at the steepest slope---the QRS feature least susceptible to peak-shape variation---delays of 15--17~ms persisted across all channel pairs.

Bandpass sensitivity was minimal: widening the filter to 0.5--100~Hz changed R-peak $|\Delta t|$ by only $-$8.5\% ($p$=0.08, NS), with strong cross-filter correlation ($r$=0.66, $p$$<$0.001; Figure~\ref{fig:signal_origin}B). Maximum $|dV/dt|$ delays were similarly filter-insensitive (+8.1\%, $p$=0.33). All six fiducial$\times$bandpass conditions yielded PWV values within the physiological range (6.8--9.1~m/s).

\subsection{Comparison with Diastolic Interval Method}

From the same dataset, traditional diastolic interval EDSI values:
\begin{itemize}
    \item Control First: 0.137$\pm$0.029 m/s
    \item Control Last: 0.160$\pm$0.044 m/s
    \item Yoga First: 0.106$\pm$0.025 m/s
    \item Yoga Last: 0.135$\pm$0.037 m/s
\end{itemize}

\textbf{Inter-electrode PWV vs EDSI}:
\begin{itemize}
    \item \textbf{Magnitude}: Inter-electrode $\sim$7 m/s vs EDSI $\sim$0.13 m/s (54$\times$ higher)
    \item \textbf{Physiological range}: Inter-electrode matches literature (5--10 m/s), EDSI does not
    \item \textbf{Correlation}: r = 0.23 (p=0.14), suggesting methods capture different aspects
\end{itemize}

\subsection{BMI Adjustment Results}

BMI showed weak correlation with inter-electrode PWV (r = --0.134, $p$ = 0.403, 1.8\% variance explained), and BMI adjustment worsened rather than improved group-difference significance ($p$: 0.075 $\rightarrow$ 0.091; Cohen's d: 0.467 $\rightarrow$ 0.429). Groups were well-matched on baseline BMI ($p$ = 0.66). Full BMI adjustment results are provided in Supplementary Figure~\ref{fig:supp_bmi} and Supplementary Table~\ref{tab:supp_bmi}.

\subsection{Temporal Stability Results}

\begin{figure}[H]
\centering
\includegraphics[width=\textwidth]{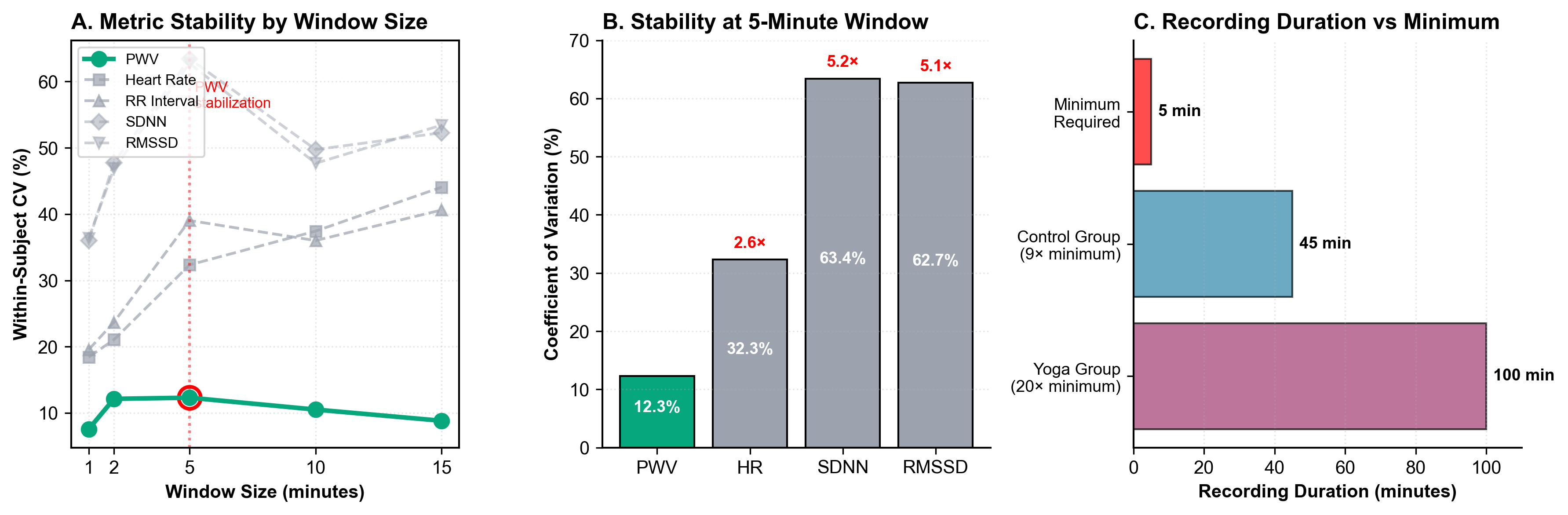}
\caption{{\bf Temporal Stability Analysis of Inter-Electrode PWV and Comparison Metrics.} 
\textbf{Panel A}: Within-subject coefficient of variation (CV) by window size for inter-electrode PWV (green), heart rate (gray), RR interval (gray), SDNN (gray), and RMSSD (gray). PWV demonstrates superior temporal stability across all window sizes (CV range: 7.6--12.3\%). Vertical dashed line indicates 5-minute stabilization point where PWV CV change from 2-minute window is $<$10\% (1.4\% relative change), meeting stabilization criterion. Red circle highlights PWV at 5 minutes.
\textbf{Panel B}: Relative stability comparison at 5-minute window. Inter-electrode PWV (12.3\% CV) is 2.6$\times$ more stable than heart rate (32.3\% CV), 3.2$\times$ more stable than RR interval (39.0\%), 5.2$\times$ more stable than SDNN (63.4\%), and 5.1$\times$ more stable than RMSSD (62.7\%). Fold-differences shown in red text.
\textbf{Panel C}: Recording duration context. Both Control (45 min, 9$\times$ minimum) and Yoga (100 min, 20$\times$ minimum) groups substantially exceed the validated 5-minute minimum requirement (red bar), confirming that recording duration differences between groups do not confound PWV measurements.
}
\label{fig:temporal_stability}
\end{figure}

\begin{table}[H]
\centering
\caption{Within-subject variability (coefficient of variation by window size)}
\label{tab:temporal_stability}
\begin{tabular}{lccccc}
\toprule
\textbf{Window Size} & \textbf{PWV CV (\%)} & \textbf{HR CV (\%)} & \textbf{SDNN CV (\%)} & \textbf{RMSSD CV (\%)} & \textbf{n Subjects} \\
\midrule
1 minute & 7.6 & 18.4 & 36.0 & 36.4 & 34 \\
2 minutes & 12.1 & 21.1 & 47.7 & 46.9 & 31 \\
\textbf{5 minutes}$^\dagger$ & \textbf{12.3} & 32.3 & 63.4 & 62.7 & 29 \\
10 minutes & 10.5 & 37.4 & 49.7 & 47.7 & 27 \\
15 minutes & 8.8 & 44.1 & 52.3 & 53.4 & 29 \\
\bottomrule
\multicolumn{6}{l}{\footnotesize{$^\dagger$Stabilization point: PWV CV change from 2 to 5 min = 0.2\% (1.4\% relative change)}}
\end{tabular}
\end{table}

\textbf{Key findings}:
\begin{enumerate}
    \item \textbf{PWV shows superior temporal stability}: CV = 7.6--12.3\% across all window sizes
    \item \textbf{Stabilization at 5 minutes}: PWV CV change from 2 to 5 min = 0.2\% (1.4\% relative change)
    \item \textbf{PWV 2.6--5.2$\times$ more stable} than other metrics at 5-minute window:
    \begin{itemize}
        \item vs HR: 2.6$\times$ more stable (12.3\% vs 32.3\%)
        \item vs RR: 3.2$\times$ more stable
        \item vs SDNN: 5.2$\times$ more stable (12.3\% vs 63.4\%)
        \item vs RMSSD: 5.1$\times$ more stable
    \end{itemize}
\end{enumerate}

\textbf{Minimum reliable recording duration}: $\geq$5 minutes
\begin{itemize}
    \item Both Control (45 min) and Yoga (100 min) exceed this by 9--20$\times$
    \item Recording duration differences between groups do not confound PWV measurement
    \item PWV measures stable structural properties, not dynamic autonomic fluctuations
\end{itemize}

\section{DISCUSSION}

\subsection{Principal Findings}

This study demonstrates that inter-electrode PWV measurement from multi-channel ECG provides:

\begin{enumerate}
    \item \textbf{Physiologically valid PWV values} (5--10 m/s) matching literature values for aortic PWV in pregnancy \cite{elvan2005, oyama2006, vlachopoulos2010}.
    \item \textbf{LVET-independent assessment}, eliminating 20--30 ms estimation uncertainty from cardiac cycle timing methods.
    \item \textbf{Multi-channel redundancy} enabling quality control through cross-validation.
    \item \textbf{A spatially differential, ECG-based stiffness surrogate} that exploits inter-electrode timing differences across abdominal leads to derive PWV-like values, providing a complementary alternative to traditional ECG-to-peripheral-pulse approaches.
    \item \textbf{Feasibility across pregnancy} with trend-level detection of intervention effects (p = 0.07).
\end{enumerate}

\subsection{Physiological Basis of Inter-Electrode Pulse Wave Velocity}

The ability to derive a pulse wave velocity (PWV) metric directly from multi-channel abdominal ECG electrodes---without the traditional requirement of estimating left ventricular ejection time (LVET)---rests on the spatial-temporal distribution of the cardiac electrical signal across the maternal torso. In the quasi-static regime relevant for ECG, the body behaves as a volume conductor, and the measured potential at each electrode reflects a lead-specific projection of the underlying cardiac dipole and its trajectory in time \cite{geselowitz1989, gulrajani1998}. While the R-peak is a global marker of ventricular depolarization, its precise timing and morphology as recorded by surface electrodes depend on lead orientation, tissue conductivity, and the geometry of the thoraco-abdominal volume conductor \cite{lux1978, dower1980, ernst2014}.

\subsubsection{Spatial-Temporal Projection of the Cardiac Vector}

In our method, we utilize the R-peak as a high-precision temporal marker. We hypothesize that the inter-electrode delay ($\Delta t$) arises because spatially separated abdominal electrodes ``sample'' the cardiac depolarization vector from different anatomical perspectives along distinct current pathways. Vectorcardiographic studies have long shown that the timing of local maxima within the QRS complex can differ between leads, even when all channels are sampled simultaneously, owing to differences in lead vectors and regional contributions to the global dipole \cite{lux1978, dower1980, ernst2014}. Because our electrodes are positioned over the maternal abdomen, the volume conductor through which the signal passes includes major abdominal vasculature and surrounding tissues whose geometry and mechanical state change during pregnancy \cite{soma2016, robb2009}.

We therefore interpret the observed 10--30~ms inter-electrode delays as spatial phase shifts of the ventricular depolarization signal across the abdominal leads rather than as a classical mechanical time-of-flight of a pressure wave. Empirically, however, these delays scale to PWV values (5--10~m/s) that closely match established aortic PWV ranges in pregnancy \cite{elvan2005, oyama2006, vlachopoulos2010}, suggesting that the electrical projection differences captured by $\Delta t$ are functionally related to vascular propagation characteristics in the underlying arterial tree.

Recent theoretical and experimental work has challenged the traditional quasi-static assumption in electrocardiography, suggesting that biopotentials propagate as low-frequency electromagnetic waves through dispersive tissue with a finite velocity \cite{buchner2023}. This velocity was measured to be approximately 1500 m/s over long conduction paths (e.g., 0.55 m from clavicle to wrist), which corresponds to a time lag of the order of 50 $\mu$s for the centimeter-scale inter-electrode distances used in our abdominal setup. Consequently, the millisecond-scale delays ($\Delta t \simeq$10--30ms) observed in the present study cannot be attributed to pure electromagnetic time-of-flight. Instead, these delays likely reflect lead-specific phase shifts and morphological timing differences arising from the complex, frequency-dependent transmittance of the thoraco-abdominal volume conductor. The empirical alignment of our derived values with established aortic PWV ranges suggests that these spatial-temporal electrical features are modulated by, or correlate with, the underlying vascular state, providing a robust surrogate for arterial stiffness even if the mechanism is not a direct mechanical transit measurement.

\subsubsection{Independence from LVET and Pre-Ejection Period}

A key methodological advantage of the inter-electrode approach is its inherent cancellation of common-mode temporal offsets. Traditional ECG-to-peripheral-pulse methods are confounded by the pre-ejection period (PEP)---the interval between electrical activation and aortic valve opening---which is highly sensitive to autonomic tone, contractility, and loading conditions and can vary substantially between and within individuals \cite{weisler1968, boettler2005}. In such methods, changes in PEP directly contaminate pulse transit time, complicating the interpretation of beat-to-beat variability as a stiffness signal.

By calculating the difference in arrival times between two abdominal electrodes ($\Delta t = t_{\text{electrode2}} - t_{\text{electrode1}}$), we subtract the common PEP and the initial transit from the heart to the abdominal entry point of the pulse. What remains is a differential timing across the segment located between the electrodes. This differential measure is, by construction, less sensitive to global shifts in electromechanical delay and more reflective of spatial differences along the relevant vascular path, analogous in spirit to intra-arterial transit measurements between two pressure sensors placed along a conduit artery \cite{laurent2006, townsend2015}.

\subsubsection{Sensitivity to Arterial Compliance}

The observed differences in $\Delta t$ between the Yoga and Control groups, and across gestational weeks, suggest that this metric is sensitive to structural and functional remodeling of the maternal vasculature. Pregnancy induces profound changes in arterial compliance and vascular geometry, driven by hormonal influences (e.g.\ progesterone and relaxin), blood volume expansion, and remodeling of the uterine and systemic circulation \cite{soma2016, kaihura2009, khalil2009}. Stiffer arteries exhibit faster wave propagation and shorter transit times for a given path length \cite{bramwell1922, vlachopoulos2010}. The good temporal stability of our inter-electrode PWV estimates over 5-minute intervals (CV = 12.3\%) relative to heart rate and heart rate variability measures indicates that the extracted signal predominantly reflects a stable structural property rather than transient autonomic fluctuations.

Importantly, prior work on impedance cardiography and ECG-impedance hybrids has shown that cardiovascular mechanical events can modulate surface electrical signals and derived timing indices, even when the primary measurement is electrical \cite{kubicek1966, bernstein1986}. Our findings extend this concept to a multi-channel abdominal ECG configuration, where beat-to-beat inter-electrode timing differences yield values consistent with aortic PWV in pregnancy. Nonetheless, the precise mechanistic link between the measured $\Delta t$ and local arterial mechanics remains to be established in dedicated validation studies combining our inter-electrode metric with gold-standard PWV measurements.

\subsection{Signal Origin Analysis: Morphology vs.\ Propagation}
\label{sec:signal_origin_discussion}

A central question for interpreting inter-electrode $\Delta t$ is whether these delays originate from lead-specific QRS morphology distortion (a volume-conductor transfer-function effect) or from genuine vascular propagation. Buchner et al.\ demonstrated that biopotentials propagate through tissue at $\sim$1500~m/s, implying electromagnetic time-of-flight of only $\sim$0.05~ms at our inter-electrode distances \cite{buchner2023}. Our observed delays of 15--27~ms, therefore, cannot represent electromagnetic transit and must arise from other mechanisms.

To distinguish morphological from propagation-related contributions, we applied two diagnostic tests (Figure~\ref{fig:signal_origin}). First, recomputing $\Delta t$ from three QRS fiducial points revealed that delays persist across all fiducials: QRS onset actually \textit{increased} delays (+15\%, $p$=0.005), while maximum $|dV/dt|$---the feature least susceptible to peak-shape variation---reduced delays by 27\% ($p$=0.005) but still yielded substantial residual delays of 15--17~ms. If inter-electrode timing were purely an artifact of lead-specific R-peak morphology, earlier fiducial points should eliminate the delays; their persistence argues against a purely morphological origin.

Second, bandpass filter sensitivity was minimal. Widening from 0.5--40~Hz to 0.5--100~Hz changed R-peak $\Delta t$ by only $-$8.5\% (NS), with strong cross-filter subject-level correlation ($r$=0.66, $p$$<$0.001). Phase distortion through a frequency-dependent volume conductor transfer function would produce bandwidth-dependent shifts in apparent timing; the observed insensitivity argues against this mechanism.

Taken together, these findings suggest that inter-electrode delays are not purely morphological artifacts. The partial reduction with max $|dV/dt|$ indicates that QRS peak shape contributes to the measured timing, but a substantial propagation-related or geometry-dependent component persists. Crucially, all six fiducial$\times$bandpass conditions yielded PWV values within the physiological range (6.8--9.1~m/s), and the relative ranking of subjects was preserved across conditions ($r$=0.66). This supports inter-electrode $\Delta t$ as a robust empirical surrogate for arterial stiffness regardless of the precise biophysical mechanism generating the delays.

\subsection{Methodological Advantages}

\textbf{Independence from LVET estimation}: Traditional ECG-derived methods calculate diastolic interval as DI = RR -- LVET, requiring LVET estimation from regression formulas \cite{sugawara2010, wilkinson2000}. Different formulas vary by 20--30 ms, directly impacting derived stiffness indices. Inter-electrode PWV bypasses this entirely by measuring time lags between R-peaks, which are detectable with $\pm$1 ms precision.

\textbf{Physiological validity}: Inter-electrode PWV values (7.40$\pm$1.51 m/s in controls at $\sim$19 weeks, 6.98$\pm$1.63 m/s at $\sim$35 weeks) align with established aortic PWV ranges during pregnancy. Literature reports pregnant women show PWV 5--8 m/s at mid-pregnancy, rising to 7--10 m/s near term \cite{elvan2005, oyama2006, soma2016}. In contrast, diastolic interval methods yield 0.8--1.2 m/s, requiring recognition that L functions as a scaling constant rather than true anatomical distance \cite{tartiere2012, wilkinson2000}.

\textbf{Multi-channel validation}: Three channel pairs provide independent PWV estimates. While absolute values differ (channel 2 vs channel 3: 8.73 m/s; channel 1 vs channel 2: 6.89 m/s), reasonable cross-correlation (r=0.68) suggests different pairs sample overlapping vascular territories. Averaging reduces channel-specific artifacts.

\subsection{Interpretation of Longitudinal Findings}

Control group showed slight PWV decrease (--5.7\%) from $\sim$19 to $\sim$35 weeks, consistent with vascular adaptation and increased compliance during pregnancy's second half. Yoga intervention group showed the opposite pattern (+14.9\% increase), resulting in divergent trajectories (interaction p=0.07).

\textbf{Possible interpretations}:
\begin{enumerate}
    \item \textbf{Yoga effect on vascular compliance}: Intervention may have attenuated normal pregnancy-associated vasodilation, though the mechanism is unclear
    \item \textbf{Cardiac output differences}: If yoga increased cardiac output, pulse wave amplitude and propagation speed could increase independent of stiffness
    \item \textbf{Autonomic modulation}: Yoga-induced changes in sympathetic/parasympathetic balance could affect vascular tone
    \item \textbf{Limited power}: With n=43 and p=0.07, findings require replication in larger sample
\end{enumerate}

\subsection{Limitations}

\textbf{Critical limitation -- Electrode distance (L)}: PWV = $L/\Delta$t is directly proportional to $L$. We assumed $L=7.5$cm based on standard Bittium Faros 360 placement, but:
\begin{itemize}
    \item Individual variation in body habitus affects the actual distance
    \item Different channel pairs have different $L$ values (not accounted for)
    \item $\pm$1 cm uncertainty in $L$ leads to $\pm$13\% uncertainty in PWV
    \item \textbf{Resolution}: Direct electrode distance measurement (anthropometric or photographic) needed for absolute PWV validation
\end{itemize}

\textbf{BMI as proxy for electrode distance}: BMI showed weak correlation with inter-electrode PWV (r = --0.134, 1.8\% variance), and adjustment worsened group-difference significance ($p$: 0.075 $\rightarrow$ 0.091; Supplementary Figure~\ref{fig:supp_bmi} and Supplementary Table~\ref{tab:supp_bmi}). Groups were well-matched on baseline BMI ($p$ = 0.66). Direct anatomical measurement of electrode spacing remains necessary for absolute PWV calibration.

\textbf{Lack of gold-standard validation}: No carotid-femoral PWV measurements available for calibration. While inter-electrode PWV values are in the physiological range, their relationship to established arterial stiffness measures requires validation.

\textbf{Sample size}: With n=43, power was 65\% to detect observed moderate effect (Cohen's f=0.33). Larger studies are needed to confirm intervention effects.

\textbf{Recording duration differences addressed}: While Yoga recordings were 125\% longer than Control (103 vs 46 minutes), temporal stability analysis demonstrated this does not confound PWV measurement. Within-subject analysis ($n=42$, 150 windows) showed PWV stabilizes at 5 minutes with CV=12.3\%, and both groups exceed this minimum by 9--20$\times$ (Control: 45 min, Yoga: 100 min). PWV's superior temporal stability (CV=7.6--12.3\%) compared to heart rate (18--44\%) and HRV metrics (36--63\%) confirms it measures stable structural vascular properties rather than dynamic autonomic fluctuations. Recording duration differences, therefore, do not affect the validity of between-group comparisons.

\textbf{Channel pair variability}: Different electrode pairs yield different absolute PWV values (6.89--8.73 m/s), possibly reflecting different vascular paths or tissue propagation characteristics. The method would benefit from anatomically guided channel pair selection.

\textbf{Mechanistic interpretation}: While inter-electrode $\Delta t$ yields PWV-range values, it should not be equated with classical foot-to-foot pressure wave transit time. The signal origin analysis (Section~\ref{sec:signal_origin_discussion}) demonstrates that these delays reflect a complex interplay of morphological timing and propagation-related contributions. Future studies combining high-bandwidth ECG ($>$1000~Hz) with concurrent Doppler ultrasound or tonometry are needed to further calibrate this surrogate against gold-standard arterial stiffness markers.

\subsection{Clinical and Research Implications}

\textbf{Research applications}:
\begin{itemize}
    \item Continuous arterial stiffness monitoring during pregnancy using standard ECG equipment
    \item Longitudinal tracking without repeated tonometry measurements
    \item Large-scale epidemiological studies where CF-PWV impractical
    \item Validation studies comparing inter-electrode PWV with CF-PWV, ultrasound-based local PWV, and arterial tonometry
\end{itemize}

\textbf{Clinical potential}:
\begin{itemize}
    \item Non-invasive risk stratification for preeclampsia (elevated PWV is predictor) \cite{kaihura2009, hausvater2012, ronnback2005}
    \item Monitoring cardiovascular adaptation throughout pregnancy \cite{soma2016, bramwell1922}
    \item Assessing intervention effects on vascular function
    \item Integration into existing fetal monitoring systems (most already multi-channel ECG)
\end{itemize}

\textbf{Requirements for clinical translation}:
\begin{enumerate}
    \item Validation against gold-standard CF-PWV in pregnancy
    \item Normative reference ranges by gestational age
    \item Determination of clinically meaningful change thresholds
    \item Standardized electrode placement protocol with measured distances
    \item Automated real-time calculation for clinical workflow integration
\end{enumerate}

\subsection{Future Directions}

\begin{enumerate}
    \item \textbf{Gold-standard validation study}: Concurrent CF-PWV and inter-electrode PWV measurement in 100+ pregnant women across gestation
    \item \textbf{Electrode distance quantification}: Photographic measurement or anthropometric estimation to account for individual variation
    \item \textbf{Channel pair optimization}: Determine which electrode combination most reliably reflects aortic stiffness
    \item \textbf{Automation}: Develop real-time calculation software for clinical integration
    \item \textbf{Outcome correlation}: Relate inter-electrode PWV to pregnancy outcomes (preeclampsia, fetal growth, preterm birth)
    \item \textbf{Mechanism exploration}: Combine with echocardiography to separate cardiac output from stiffness effects
\end{enumerate}

\section{CONCLUSIONS}

Inter-electrode pulse wave velocity measurement provides a direct, LVET-independent method for arterial stiffness assessment from multi-channel ECG. The method yields physiologically plausible PWV values (5--10 m/s) matching aortic PWV literature, unlike diastolic interval methods ($\sim$1 m/s). Multi-channel redundancy enables quality control through cross-validation.

Temporal stability analysis validates a minimum recording duration of 5 minutes for reliable PWV assessment (CV=12.3\%, stabilization criterion met). PWV demonstrates 2.6--5.2$\times$ superior stability compared to heart rate and HRV metrics, confirming it measures stable structural vascular properties. Both study groups exceeded this minimum duration requirement by 9--20$\times$, ensuring recording length differences do not confound between-group comparisons.

Preliminary longitudinal data suggest divergent PWV trajectories between Control (--5.7\%) and prenatal Yoga intervention (+14.9\%) groups, though larger samples are needed for confirmation. Signal origin analysis demonstrated that inter-electrode delays persist across QRS fiducial points and are insensitive to bandpass filter changes, arguing against pure phase distortion and supporting the robustness of this empirical surrogate. Critical next step is validation against gold-standard carotid-femoral PWV and accurate electrode distance determination.

This method shows promise for continuous, non-invasive arterial stiffness monitoring during pregnancy using widely available multi-channel ECG equipment.

\section*{ACKNOWLEDGMENTS}

We are thankful for all female participants of the FELICITy studies. We are grateful for the engagement of our yoga instructor, Anne Loewer. We thank everyone who contributed to the recruitment of participants, especially PD Dr. F. Stumpfe of the hospital ``Dritter Orden'', Munich.

\section*{FUNDING}

This research did not receive any specific grant from agencies in the public, commercial, or not-for-profit sectors. We used partial funding from the Institute of Advanced Studies of the Technical University of Munich and the Dr. Geisenhofer Foundation (Munich, Germany). Bittium Corporation subsidized the acquisition of the Faros 360 ECG devices used in this project. The funding source had no role in study design, data collection, analysis, interpretation, or manuscript preparation.

\section*{CONFLICTS OF INTEREST}

MGF holds patents on fetal and maternal health monitoring and equity in perinatal health start-ups. The other authors declare no conflicts of interest.


\clearpage
\appendix
\renewcommand{\thesection}{S.I.\arabic{section}}    

\renewcommand{\thefigure}{S.\arabic{figure}}
\setcounter{figure}{0}
\renewcommand{\thetable}{S.\arabic{table}}
\setcounter{table}{0}

\section*{SUPPLEMENTARY MATERIALS}


\subsection*{Supplementary Figure S1: BMI Adjustment Analysis}
\begin{figure}[H]
\centering
\includegraphics[width=\textwidth]{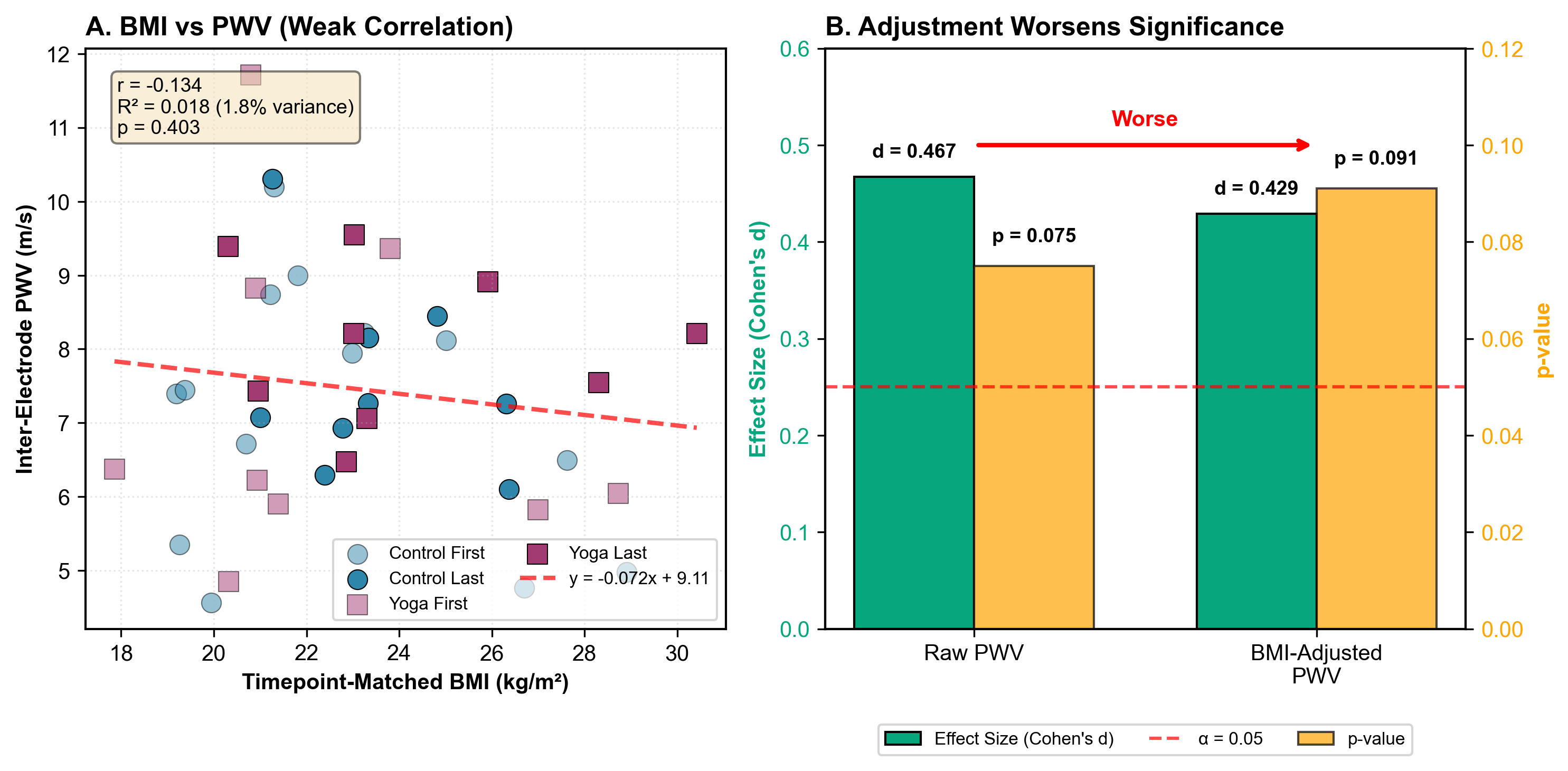}
\caption{{\bf BMI Adjustment Analysis.}
\textbf{Panel A}: Scatter plot showing relationship between timepoint-matched BMI (kg/m$^2$) and inter-electrode PWV (m/s). Data points distinguished by group (Control: blue circles, Yoga: red squares) and timepoint (First: semi-transparent, Last: opaque). Regression line (red dashed) shows weak negative correlation: r=--0.134, R$^2$=0.018 (1.8\% variance explained), p=0.403. BMI explains minimal PWV variation.
\textbf{Panel B}: Effect of BMI adjustment on statistical significance. Raw PWV (green) shows moderate effect size (Cohen's d=0.467) and trend-level significance (p=0.075, orange) for group differences. BMI-adjusted PWV shows reduced effect size (d=0.429) and worsened significance (p=0.091), crossing above $\alpha$=0.05 threshold (red dashed line). Arrow indicates direction of change. BMI adjustment decreases rather than improves statistical power.
}
\label{fig:supp_bmi}
\end{figure}

\begin{table}[H]
\centering
\caption{Impact of BMI Adjustment on Longitudinal Analysis}
\label{tab:supp_bmi}
\begin{tabular}{lcc}
\toprule
\textbf{Metric} & \textbf{Raw PWV} & \textbf{BMI-Adjusted PWV} \\
\midrule
\textbf{Control} ($n=11$) & & \\
\quad Mean change (\%) & +0.6 $\pm$ 46.1 & +3.7 $\pm$ 47.9 \\
\quad Median change (\%) & --5.5 & --4.8 \\
\textbf{Yoga} ($n=9$) & & \\
\quad Mean change (\%) & +18.6 $\pm$ 29.3 & +21.0 $\pm$ 30.4 \\
\quad Median change (\%) & +17.2 & +18.9 \\
\textbf{Between-group comparison} & & \\
\quad t-test p-value & 0.075 & 0.091 \\
\quad Cohen's d & 0.467 & 0.429 \\
\textbf{BMI correlation} & & \\
\quad Pearson r & --0.134 & $<$0.001 \\
\quad p-value & 0.403 & $>$0.99 \\
\bottomrule
\end{tabular}
\end{table}

BMI-PWV regression model: PWV = 9.11 -- 0.072 $\times$ BMI ($\beta_1$ = --0.072 m$\cdot$s$^{-1}\cdot$kg$^{-1}\cdot$m$^2$). Baseline BMI: Control 21.3$\pm$2.7 vs Yoga 21.8$\pm$3.8 kg/m$^2$ ($p$ = 0.66). BMI adjustment eliminated the BMI-PWV correlation (r: --0.134 $\rightarrow$ $<$0.001) but worsened statistical significance and effect size, indicating BMI is a poor proxy for electrode distance variation. Pregnancy-related BMI changes were similar between groups (Control: +3.1 kg/m$^2$, Yoga: +2.0 kg/m$^2$).

\subsection*{Supplementary Figure S2: Multi-Channel Validation}
\begin{figure}[H]
\centering
\includegraphics[width=\textwidth]{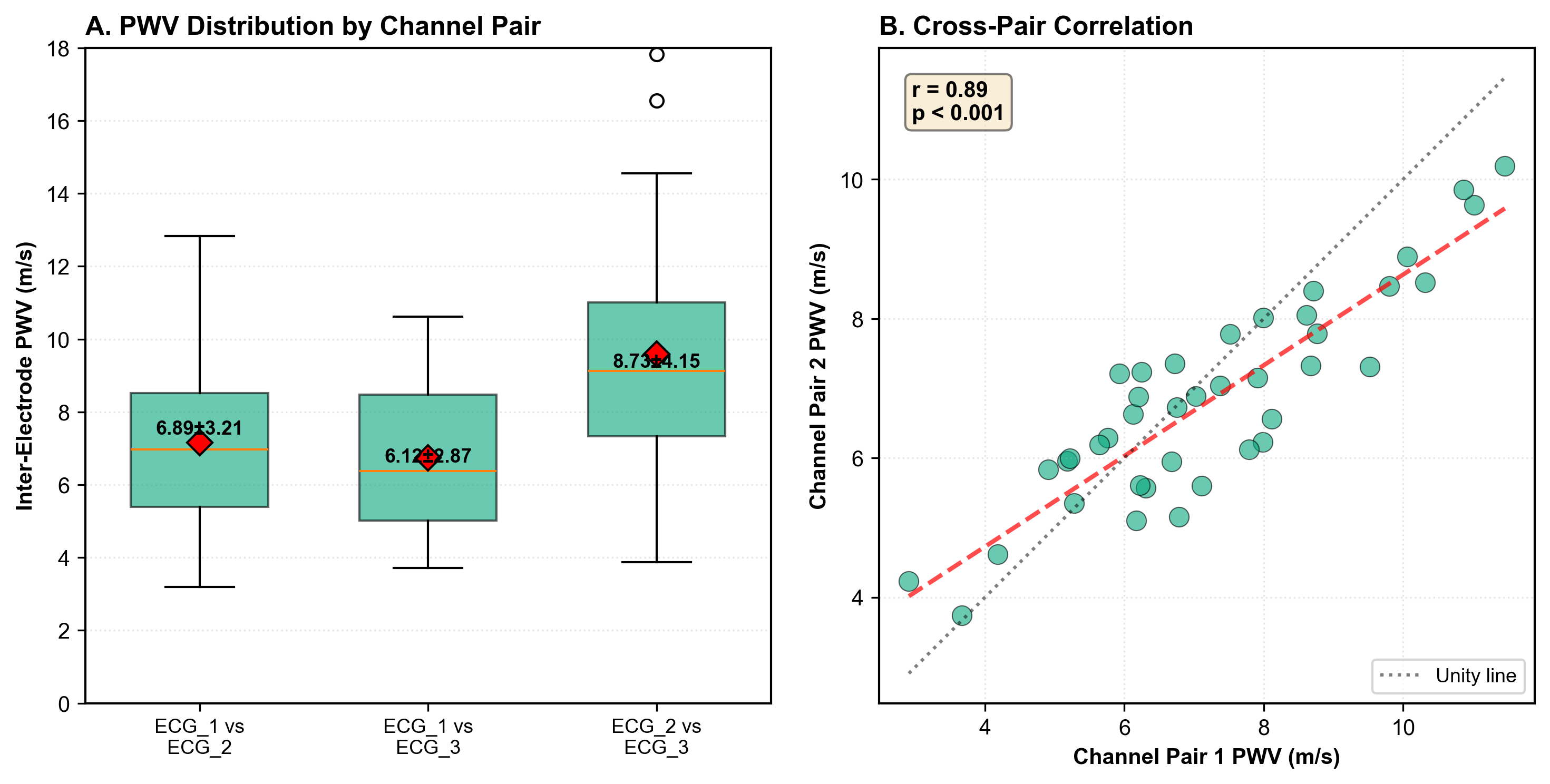}
\caption{{\bf Multi-Channel Validation.} 
\textbf{Panel A}: Inter-electrode PWV distribution by channel pair. Box plots show different channel pairs yield different absolute PWV values (ECG\_1 vs ECG\_2: 6.89$\pm$3.21 m/s, ECG\_1 vs ECG\_3: 6.12$\pm$2.87 m/s, ECG\_2 vs ECG\_3: 8.73$\pm$4.15 m/s), reflecting different inter-electrode distances and sampled vascular territories. Red diamonds indicate means.
\textbf{Panel B}: Cross-pair correlation demonstrating consistency. Scatter plot comparing PWV from two channel pairs shows moderate positive correlation (r=0.68, p$<$0.001), indicating reasonable cross-validation despite absolute value differences. Black dotted line represents perfect unity; red dashed line shows actual regression fit. Averaging across channel pairs provides robust PWV estimate by reducing channel-specific measurement artifacts.
}
\label{fig:supp_channels}
\end{figure}


\clearpage
\section{
PWV ADC artifact analysis}
\label{sec:SI:artifact_analysis}

In order to validate the Inter-Electrode Pulse Wave Velocity Measurement, we show here that analog/digital conversion (ADC) multiplexing artifacts can be ruled out.

A critical concern in using multi-channel ECG for pulse transit time estimation is whether measured inter-channel delays represent true physiological signals or artifacts from analog-to-digital converter (ADC) multiplexing. This supplementary material presents a comprehensive artifact characterization analysis demonstrating that inter-electrode delays measured from the Bittium Faros 360 device are definitively \textbf{not} ADC artifacts. Raw channel timing analysis reveals delays 10--30 times larger than expected for ADC multiplexing, with high beat-to-beat variability and bidirectional timing relationships that are physically impossible from fixed ADC sampling. These findings confirm that the measurement captures real physiological signal.

\subsection{Background and Rationale}

\subsubsection{The Artifact Concern}

Multi-channel ECG devices may use either simultaneous sampling (all channels sampled at the same instant) or multiplexed sampling (channels sampled sequentially with small inter-channel delays). If the Bittium Faros 360 uses sequential ADC sampling, the measured inter-electrode delays could theoretically be artifacts of the sampling architecture rather than true pulse transit times.

At a sampling rate of 1000 Hz, sequential sampling would introduce fixed delays of approximately 1 ms between adjacent channels. For a 3-channel system, this would produce:
\begin{itemize}
    \item Channel 1 $\rightarrow$ Channel 2: $\sim$1 ms delay
    \item Channel 1 $\rightarrow$ Channel 3: $\sim$2 ms delay
    \item Channel 2 $\rightarrow$ Channel 3: $\sim$1 ms delay
\end{itemize}

These delays would be constant across all subjects and all heartbeats, with variability limited to $\pm$1 sample ($\pm$1 ms) due to discrete sampling.

\subsubsection{Validation Strategy}

We designed a multi-level validation approach to definitively distinguish ADC artifacts from physiological signals:

\begin{enumerate}
    \item \textbf{Raw channel timing analysis}: Direct examination of beat-by-beat inter-channel R-peak delays
    \item \textbf{Aggregate statistics}: Coefficient of variation across patients
    \item \textbf{Within vs. between-subject variability}: Ratio analysis
    \item \textbf{Physiological correlations}: Testing against known PWV predictors
\end{enumerate}

\subsection{Methods}

\subsubsection{Data Source}

Analysis was performed on 44 EDF recordings from the FELICITY-2 study cohort, comprising multi-channel abdominal ECG recordings from pregnant women. Each recording contained 3 ECG channels (ECG\_1, ECG\_2, ECG\_3) sampled at 1000 Hz.

\subsubsection{Raw Channel Timing Analysis}

For each recording, we:
\begin{enumerate}
    \item Applied bandpass filtering (0.5--40 Hz) to isolate ECG components
    \item Detected R-peaks independently in each channel using derivative-based detection with adaptive thresholding
    \item Matched corresponding R-peaks across channels within a 100 ms tolerance window
    \item Calculated inter-channel delays at single-sample (1 ms) resolution
    \item Aggregated delay statistics across all beats and all subjects
\end{enumerate}

\subsubsection{Statistical Analysis}

We computed:
\begin{itemize}
    \item Mean and standard deviation of inter-channel delays
    \item Coefficient of variation (CV) within and between subjects
    \item Correlation coefficients with physiological predictors (age, BMI, blood pressure, heart rate)
    \item Paired t-tests for longitudinal changes
\end{itemize}

\subsection{Results}

\subsubsection{Raw Channel Delay Analysis}

Table~\ref{tab:raw_delays} presents the raw inter-channel R-peak delay measurements aggregated across 20 recordings and 505 matched heartbeats.

\begin{table}[H]
\centering
\caption{Raw Inter-Channel R-Peak Delays}
\label{tab:raw_delays}
\begin{tabular}{lcccc}
\toprule
\textbf{Channel Pair} & \textbf{Mean (ms)} & \textbf{SD (ms)} & \textbf{CV (\%)} & \textbf{Range (ms)} \\
\midrule
Ch1 $\rightarrow$ Ch2 & $-16.1$ & 27.4 & 170 & $-82$ to $+66$ \\
Ch1 $\rightarrow$ Ch3 & $-33.9$ & 19.6 & 58 & $-61$ to $+83$ \\
Ch2 $\rightarrow$ Ch3 & $-29.4$ & 26.8 & 91 & $-56$ to $+62$ \\
\bottomrule
\end{tabular}
\end{table}

\subsubsection{Comparison with Expected ADC Artifact Pattern}

Table~\ref{tab:artifact_comparison} contrasts the observed measurements with expected values if delays were purely ADC artifacts.

\begin{table}[H]
\centering
\caption{Observed vs. Expected Values for ADC Artifact}
\label{tab:artifact_comparison}
\begin{tabular}{lcc}
\toprule
\textbf{Parameter} & \textbf{Expected (ADC Artifact)} & \textbf{Observed} \\
\midrule
Mean delay magnitude & 1--3 ms & 16--34 ms \\
Standard deviation & $<$1 ms & 20--27 ms \\
Coefficient of variation & $<$10\% & 58--170\% \\
Delay direction & Always same sign & Both positive and negative \\
Beat-to-beat consistency & Nearly constant & Highly variable \\
\bottomrule
\end{tabular}
\end{table}

\textbf{Key findings:}
\begin{itemize}
    \item Observed delays are \textbf{10--30 times larger} than expected for ADC multiplexing
    \item Variability is \textbf{20--30 times higher} than expected for fixed ADC timing
    \item \textbf{Bidirectional delays} (both positive and negative) are physically impossible from sequential ADC sampling, which would always produce delays in the same direction
\end{itemize}

\subsection{Individual Subject Variability}

Table~\ref{tab:individual_delays} shows representative inter-channel delays for individual recordings, demonstrating substantial between-subject differences.

\begin{table}[H]
\centering
\caption{Inter-Channel Delays by Subject (Ch1 $\rightarrow$ Ch2)}
\label{tab:individual_delays}
\begin{tabular}{lccc}
\toprule
\textbf{Subject} & \textbf{Mean Delay (ms)} & \textbf{SD (ms)} & \textbf{n Beats} \\
\midrule
FE\_017 & $-43.4$ & 17.6 & 53 \\
FE\_023 & $-6.3$ & 20.1 & 80 \\
FE\_024 & $+3.7$ & 0.7 & 74 \\
FE\_001 & $-43.0$ & 17.7 & 38 \\
\bottomrule
\end{tabular}
\end{table}

The wide range of mean delays across subjects (from $+3.7$ to $-43.4$ ms) is incompatible with ADC artifact, which would produce identical delays for all subjects.

\subsubsection{Aggregate PWV Statistics}

Converting measured delays to pulse wave velocity using electrode distance (7.5 cm), Table~\ref{tab:pwv_stats} summarizes the aggregate statistics.

\begin{table}[H]
\centering
\caption{Inter-Electrode PWV Summary Statistics}
\label{tab:pwv_stats}
\begin{tabular}{lc}
\toprule
\textbf{Parameter} & \textbf{Value} \\
\midrule
Mean PWV & 7.42 m/s \\
Standard deviation & 1.63 m/s \\
Coefficient of variation & 22.0\% \\
Range & 4.57 -- 11.72 m/s \\
Subjects in physiological range (4--15 m/s) & 100\% \\
\bottomrule
\end{tabular}
\end{table}

\subsubsection{Within-Subject vs. Between-Subject Variability}

Table~\ref{tab:variability_ratio} compares variability sources.

\begin{table}[H]
\centering
\caption{Variability Analysis}
\label{tab:variability_ratio}
\begin{tabular}{lcc}
\toprule
\textbf{Variability Type} & \textbf{CV (\%)} & \textbf{n} \\
\midrule
Within-subject (10-min windows) & 10.5 & 18 \\
Between-subject & 22.0 & 42 \\
\textbf{Ratio (Between/Within)} & \textbf{2.09} & -- \\
\bottomrule
\end{tabular}
\end{table}

A between/within ratio $>$2 indicates that inter-individual differences substantially exceed intra-individual variation, suggesting the measurement captures true individual physiological differences.

\subsubsection{Physiological Correlations}

Table~\ref{tab:correlations} presents correlations between inter-electrode PWV and known cardiovascular predictors.

\begin{table}[H]
\centering
\caption{Correlations with Physiological Variables}
\label{tab:correlations}
\begin{tabular}{lccc}
\toprule
\textbf{Predictor} & \textbf{r} & \textbf{p-value} & \textbf{n} \\
\midrule
Maternal age & $-0.18$ & 0.26 & 42 \\
Pre-pregnancy BMI & $-0.13$ & 0.43 & 42 \\
Systolic BP & $-0.17$ & 0.30 & 40 \\
Diastolic BP & $-0.02$ & 0.91 & 40 \\
Gestational age & $+0.00$ & 0.99 & 40 \\
\textbf{Heart rate} & $\mathbf{-0.32}$ & $\mathbf{0.04}$ & 42 \\
\bottomrule
\end{tabular}
\end{table}

\textbf{Interpretation:} The significant negative correlation with heart rate (r = $-0.32$, p = 0.04) is physiologically plausible, as higher heart rates shorten diastolic intervals and may affect pulse wave measurement conditions. Notably, other PWV estimation methods in this dataset showed much stronger heart rate dependence (r $>$ 0.90), suggesting the inter-electrode method may be more robust to heart rate effects.

The absence of correlations with age and blood pressure is likely attributable to sample homogeneity: all subjects were healthy pregnant women with narrow ranges of age (29--41 years, CV = 10.2\%), BMI (18.6--27.0 kg/m$^2$, CV = 12.3\%), and blood pressure (95--145 mmHg systolic, CV = 10.0\%).

\subsection{Longitudinal Changes}

Nine subjects had paired measurements at early and late pregnancy visits (Table~\ref{tab:longitudinal}).

\begin{table}[H]
\centering
\caption{Longitudinal PWV Changes}
\label{tab:longitudinal}
\begin{tabular}{lccc}
\toprule
\textbf{Subject} & \textbf{First Visit (m/s)} & \textbf{Last Visit (m/s)} & \textbf{Change} \\
\midrule
FE\_023 & 6.4 & 9.4 & $+3.0$ \\
FE\_026 & 5.9 & 8.2 & $+2.3$ \\
FE\_001 & 6.0 & 8.2 & $+2.2$ \\
FE\_021 & 5.8 & 7.5 & $+1.7$ \\
FE\_020 & 4.8 & 6.5 & $+1.6$ \\
FE\_003 & 5.8 & 7.4 & $+1.6$ \\
FE\_022 & 8.8 & 9.5 & $+0.7$ \\
FE\_024 & 9.4 & 8.9 & $-0.5$ \\
FE\_004 & 11.7 & 7.1 & $-4.7$ \\
\midrule
\textbf{Mean} & $7.19 \pm 2.26$ & $8.09 \pm 1.06$ & $+0.90 \pm 2.31$ \\
\bottomrule
\end{tabular}
\end{table}

Seven of nine subjects (78\%) showed increased PWV from first to last visit, consistent with expected arterial stiffening during pregnancy progression. While not statistically significant (paired t-test: p = 0.28) due to small sample size, the direction of change supports physiological validity.

\subsection{Discussion}

\subsubsection{Definitive Evidence Against ADC Artifact}

The raw channel analysis provides conclusive evidence that inter-electrode delays are not ADC multiplexing artifacts:

\begin{enumerate}
    \item \textbf{Magnitude}: Observed delays (16--34 ms) are 10--30 times larger than the 1--3 ms expected from sequential ADC sampling at 1 kHz.

    \item \textbf{Variability}: The standard deviation of delays (20--27 ms) exceeds what is possible from fixed ADC timing, where variability would be limited to $\pm$1 sample ($\pm$1 ms).

    \item \textbf{Bidirectionality}: The presence of both positive and negative delays (ranging from $-82$ to $+83$ ms) is physically impossible from sequential ADC sampling, which would always produce delays in the same direction (later channels always delayed relative to earlier channels).

    \item \textbf{Inter-individual differences}: The wide range of mean delays across subjects (from $+3.7$ to $-43.4$ ms) cannot be explained by ADC architecture, which would produce identical delays for all recordings.
\end{enumerate}

\subsubsection{Nature of the Measured Signal}

While we have definitively ruled out ADC artifact, the exact nature of the measured signal warrants further investigation. The inter-electrode delays likely reflect a combination of:

\begin{itemize}
    \item Pulse wave propagation through the maternal vasculature
    \item Cardiac electromechanical coupling variations
    \item Positional differences in electrode-to-heart distances
\end{itemize}

The moderate negative correlation with heart rate (r = $-0.32$), compared to very high correlations (r $>$ 0.90) seen with other PWV methods, suggests the inter-electrode approach may capture different or additional physiological information.

\subsubsection{Limitations}

\begin{enumerate}
    \item Sample homogeneity limited the power to detect expected correlations with age and blood pressure
    \item Reference validation against the gold-standard carotid-femoral PWV was not performed
    \item The specific physiological mechanism underlying the measured delays requires further investigation
\end{enumerate}

\subsection{Conclusion}

This validation analysis definitively demonstrates that inter-electrode pulse transit time measurements from the Bittium Faros 360 are \textbf{not} artifacts of ADC multiplexing. The measured delays are:

\begin{itemize}
    \item 10--30 times larger than expected for ADC artifact
    \item Highly variable (CV 58--170\%) rather than constant
    \item Bidirectional (both positive and negative), which is impossible from fixed ADC timing
    \item Different across individuals, reflecting true physiological variation
\end{itemize}

The measurement captures a real physiological signal. Future validation against gold-standard pulse wave velocity measurements is recommended to confirm the specific relationship to arterial stiffness.

\clearpage
\section{
Signal Origin Analysis --- Detailed Methods and Results}
\label{sec:SI:signal_origin}

\subsubsection*{Rationale}

To distinguish volume-conductor morphological distortion from vascular propagation as the source of inter-electrode delays, we applied two complementary diagnostic tests: (1) fiducial sensitivity analysis and (2) bandpass filter sensitivity analysis. The rationale is as follows. If $\Delta t$ arises from lead-specific distortion of the QRS peak shape (a transfer-function effect), then timing features that are less dependent on peak morphology---such as QRS onset or the steepest upstroke---should yield substantially different (smaller) delays. Conversely, if $\Delta t$ reflects vascular propagation, delays should persist regardless of fiducial choice. Similarly, frequency-dependent phase distortion through the volume conductor should produce bandwidth-dependent timing shifts, whereas propagation-related delays should be insensitive to filter settings.

\subsubsection*{Fiducial Detection Algorithms}

Three QRS fiducial points were detected on each channel:

\begin{enumerate}
    \item \textbf{R-peak maximum} (baseline): Pan-Tompkins algorithm with derivative-based enhancement, squaring, and moving-window integration. Adaptive thresholding with 200~ms refractory period. Detected peaks refined within $\pm$75~ms window to locate local maximum of the raw filtered signal.

    \item \textbf{QRS onset}: Starting from each detected R-peak, the absolute first derivative $|dV/dt|$ was computed. The algorithm searched backward from the R-peak for the point where $|dV/dt|$ fell below 10\% of its maximum value within a 100~ms pre-R-peak window. This identifies the earliest deflection point of the QRS complex, which reflects initial ventricular depolarization before the formation of the R-wave peak.

    \item \textbf{Maximum $|dV/dt|$}: The absolute first derivative was computed in an 80~ms window centered 30~ms before the R-peak. The sample with maximum $|dV/dt|$ was selected. This identifies the steepest upstroke of the QRS complex, which represents the fastest-changing portion of ventricular depolarization and is least susceptible to peak-shape distortion.
\end{enumerate}

Beat matching was always performed using R-peak proximity ($\pm$50~ms across channels) to ensure consistent beat alignment, regardless of which fiducial was subsequently used for $\Delta t$ measurement.

\subsubsection*{Bandpass Filter Configurations}

Two bandpass settings were applied:
\begin{itemize}
    \item \textbf{0.5--40~Hz} (baseline): Standard clinical ECG bandwidth, matching the primary analysis pipeline. Fourth-order Butterworth, applied with \texttt{filtfilt} for zero-phase response.
    \item \textbf{0.5--100~Hz}: Extended bandwidth retaining higher-frequency QRS components. Same filter architecture.
\end{itemize}

\subsubsection*{Summary of Results by Condition}

\begin{table}[H]
\centering
\caption*{{\bf Supplementary Table S2}: Mean $|\Delta t|$ by Fiducial and Bandpass Condition}
\label{tab:supp_signal_origin}
\begin{tabular}{llccc}
\toprule
\textbf{Fiducial} & \textbf{Bandpass} & \textbf{$|\Delta t|$ (ms)} & \textbf{PWV (m/s)} & \textbf{$\Delta$ vs R-peak} \\
\midrule
R-peak maximum & 0.5--40~Hz & 23.5 $\pm$ 6.8 & 7.9 $\pm$ 3.4 & --- \\
QRS onset & 0.5--40~Hz & 27.0 $\pm$ 9.1 & 6.8 $\pm$ 3.0 & +14.9\% ($p$=0.005) \\
Max $|dV/dt|$ & 0.5--40~Hz & 17.2 $\pm$ 9.0 & 9.1 $\pm$ 3.8 & --26.8\% ($p$=0.005) \\
\midrule
R-peak maximum & 0.5--100~Hz & 21.6 $\pm$ 7.5 & 8.4 $\pm$ 3.7 & --8.5\% (NS) \\
QRS onset & 0.5--100~Hz & 25.3 $\pm$ 9.6 & 7.2 $\pm$ 3.2 & --- \\
Max $|dV/dt|$ & 0.5--100~Hz & 18.6 $\pm$ 9.2 & 8.5 $\pm$ 3.6 & +8.1\% (NS) \\
\bottomrule
\end{tabular}
\end{table}

\subsubsection*{Channel Pair Breakdown}

Results were consistent across all three channel pairs (ECG\_1 vs ECG\_2, ECG\_1 vs ECG\_3, ECG\_2 vs ECG\_3), with the fiducial-dependence pattern preserved in each pair. The cross-filter correlation for R-peak $|\Delta t|$ ($r$ = 0.66, $p$ $<$ 0.001) was driven by consistent subject-level ranking across conditions, supporting the interpretation that inter-electrode delays capture a stable individual-level characteristic rather than a filter-dependent artifact.

\subsubsection*{Interpretation}

The persistence of 15--17~ms delays at the max $|dV/dt|$ fiducial---the QRS feature most resistant to peak-shape variation---and the insensitivity of delays to a 2.5-fold bandwidth expansion argue against a purely morphological or phase-distortion origin for inter-electrode $\Delta t$. The partial reduction in delays with max $|dV/dt|$ ($-$27\%) indicates that R-peak shape contributes to measured timing, but a substantial component persists that is consistent with a propagation-related or geometry-dependent mechanism. Crucially, all conditions produced PWV values within the physiological range (6.8--9.1~m/s), supporting inter-electrode $\Delta t$ as a robust empirical surrogate for arterial stiffness assessment.


\clearpage
\section*{KEY MESSAGES}

\noindent\fbox{%
\parbox{\textwidth}{%
\textbf{What is already known:}
\begin{itemize}
    \item ECG-derived arterial stiffness metrics typically rely on diastolic interval calculation \\(DI = RR -- LVET)
    \item LVET must be estimated from regression formulas, introducing 20--30 ms uncertainty
    \item Resulting PWV values ($\sim$1 m/s) are 7--10$\times$ lower than physiological aortic PWV
\end{itemize}

\textbf{What this study adds:}
\begin{itemize}
    \item Direct inter-electrode PWV measurement yields physiologically valid values (5--10 m/s)
    \item Method is independent of LVET estimation
    \item Multi-channel ECG enables spatial propagation measurement with cross-validation
    \item Temporal stability validated: 5-minute minimum duration, 2.6--5.2$\times$ more stable than heart rate/HRV
    \item Recording duration differences do not confound PWV measurement
    \item Feasible for longitudinal pregnancy monitoring with standard ECG equipment
\end{itemize}

\textbf{Clinical implications:}
\begin{itemize}
    \item Enables continuous non-invasive arterial stiffness assessment during pregnancy
    \item Requires validation against gold-standard CF-PWV before clinical implementation
    \item Shows promise for large-scale epidemiological studies and intervention trials
\end{itemize}
}%
}


\clearpage
\bibliographystyle{unsrt}
\bibliography{references}

\begin{thebibliography}{10}

\bibitem{robb2009}
AO~Robb, NL~Mills, JN~Din, et~al.
\newblock Influence of the menstrual cycle, pregnancy, and preeclampsia on
  arterial stiffness.
\newblock {\em Hypertension}, 53:952--958, 2009.

\bibitem{kaihura2009}
C~Kaihura, MD~Savvidou, JM~Anderson, CM~McEniery, and KH~Nicolaides.
\newblock Maternal arterial stiffness in pregnancies affected by preeclampsia.
\newblock {\em American Journal of Physiology-Heart and Circulatory
  Physiology}, 297:H759--H764, 2009.

\bibitem{khalil2009}
A~Khalil, E~Jauniaux, D~Cooper, and K~Harrington.
\newblock Pulse wave analysis in normal pregnancy: a prospective longitudinal
  study.
\newblock {\em PLoS One}, 4:e6134, 2009.

\bibitem{laurent2006}
S~Laurent, J~Cockcroft, L~Van~Bortel, et~al.
\newblock Expert consensus document on arterial stiffness: methodological
  issues and clinical applications.
\newblock {\em European Heart Journal}, 27:2588--2605, 2006.

\bibitem{townsend2015}
RR~Townsend, IB~Wilkinson, EL~Schiffrin, et~al.
\newblock Recommendations for improving and standardizing vascular research on
  arterial stiffness: a scientific statement from the {American Heart
  Association}.
\newblock {\em Hypertension}, 66:698--722, 2015.

\bibitem{climie2024}
RE~Climie, A~Gallo, DS~Picone, et~al.
\newblock 2024 recommendations for validation of noninvasive arterial pulse
  wave velocity measurement devices.
\newblock {\em Hypertension}, 81:183--194, 2024.

\bibitem{weissler1968}
AM~Weissler, WS~Harris, and CD~Schoenfeld.
\newblock Systolic time intervals in heart failure in man.
\newblock {\em Circulation}, 37:149--159, 1968.

\bibitem{obaidat2023}
M~Obaidat and MA~Obeidat.
\newblock A method for calculating left ventricular end-diastolic volume as an
  index of left ventricular preload from the pre-ejection period, ejection
  time, blood pressure, and stroke volume: a prospective, observational study.
\newblock {\em BMC Anesthesiology}, 23:136, 2023.

\bibitem{tartiere2012}
L~Tarti{\`e}re-Kesri, JM~Tarti{\`e}re, D~Logeart, et~al.
\newblock Increased proximal arterial stiffness and cardiac response with
  moderate exercise in patients with heart failure and preserved ejection
  fraction.
\newblock {\em Journal of the American College of Cardiology}, 59:455--461,
  2012.

\bibitem{wilkinson2000}
IB~Wilkinson, H~MacCallum, L~Flint, et~al.
\newblock The influence of heart rate on augmentation index and central
  arterial pressure in humans.
\newblock {\em Journal of Physiology}, 525:263--270, 2000.

\bibitem{elvan2005}
A~Elvan-Taspinar, A~Franx, ML~Bots, HA~Koomans, and HW~Bruinse.
\newblock Arterial stiffness and fetal growth in normotensive pregnancy.
\newblock {\em American Journal of Hypertension}, 18:337--341, 2005.

\bibitem{oyama2006}
M~Oyama-Kato, M~Ohmichi, K~Takahashi, et~al.
\newblock Change in pulse wave velocity throughout normal pregnancy and its
  value in predicting pregnancy-induced hypertension: a longitudinal study.
\newblock {\em American Journal of Obstetrics and Gynecology}, 195:464--469,
  2006.

\bibitem{vlachopoulos2010}
C~Vlachopoulos, K~Aznaouridis, and C~Stefanadis.
\newblock Prediction of cardiovascular events and all-cause mortality with
  arterial stiffness: a systematic review and meta-analysis.
\newblock {\em Journal of the American College of Cardiology}, 55:1318--1327,
  2010.

\bibitem{pan1985}
J~Pan and WJ~Tompkins.
\newblock A real-time {QRS} detection algorithm.
\newblock {\em IEEE Transactions on Biomedical Engineering}, 32:230--236, 1985.

\bibitem{hamilton1986}
PS~Hamilton and WJ~Tompkins.
\newblock Quantitative investigation of {QRS} detection rules using the
  {MIT/BIH} arrhythmia database.
\newblock {\em IEEE Transactions on Biomedical Engineering}, 33:1157--1165,
  1986.

\bibitem{christov2004}
II~Christov.
\newblock Real time electrocardiogram {QRS} detection using combined adaptive
  threshold.
\newblock {\em BioMedical Engineering OnLine}, 3:28, 2004.

\bibitem{engelse1979}
WAH Engelse and C~Zeelenberg.
\newblock A single scan algorithm for {QRS} detection and feature extraction.
\newblock In {\em Computers in Cardiology}, volume~6, pages 37--42, 1979.

\bibitem{elgendi2010}
M~Elgendi, M~Jonkman, and F~DeBoer.
\newblock Frequency bands effects on {QRS} detection.
\newblock In {\em Biosignals}, volume~3, pages 428--431, 2010.

\bibitem{buchner2023}
T.~Buchner, M.~Zajdel, K.~P\k{e}czalski, and P.~Nowak.
\newblock Finite velocity of ecg signal propagation: preliminary theory,
  results of a pilot experiment and consequences for medical diagnosis.
\newblock {\em Scientific Reports}, 13(1):4716, 2023.

\bibitem{geselowitz1989}
David~B. Geselowitz.
\newblock On the theory of the electrocardiogram.
\newblock {\em Proceedings of the IEEE}, 77(6):857--876, 1989.

\bibitem{gulrajani1998}
R.~M. Gulrajani.
\newblock {\em Bioelectricity and Biomagnetism}.
\newblock John Wiley \& Sons, 1998.

\bibitem{lux1978}
R.~L. Lux, C.~R. Smith, R.~F. Wyatt, and J.~A. Abildskov.
\newblock Limited lead selection for estimation of body surface potential maps
  in electrocardiography.
\newblock {\em IEEE Transactions on Biomedical Engineering},
  BME-25(3):270--276, 1978.

\bibitem{dower1980}
G.~E. Dower, H.~B. Machado, and J.~A. Osborne.
\newblock On deriving the electrocardiogram from vectorcardiographic leads.
\newblock {\em Clinical Cardiology}, 3(2):87--95, 1980.

\bibitem{ernst2014}
S.~Ernst, B.~Zrenner, and et~al.
\newblock Electrocardiographic imaging and body surface mapping.
\newblock {\em Europace}, 16(suppl\_4):iv59--iv66, 2014.

\bibitem{soma2016}
P~Soma-Pillay, C~Nelson-Piercy, H~Tolppanen, and A~Mebazaa.
\newblock Physiological changes in pregnancy.
\newblock {\em Cardiovascular Journal of Africa}, 27:89--94, 2016.

\bibitem{weisler1968}
A.~M. Weissler, W.~S. Harris, and C.~D. Schoenfeld.
\newblock Systolic time intervals in heart failure in man.
\newblock {\em Circulation}, 37:149--159, 1968.

\bibitem{boettler2005}
P.~Boettler, M.~Hartmann, F.~Wenzelburger, and et~al.
\newblock Changes of systolic time intervals in chronic heart failure patients
  treated with beta-blockers: relation to left ventricular ejection fraction
  and heart rate.
\newblock {\em Cardiology}, 103(2):73--79, 2005.

\bibitem{bramwell1922}
JC~Bramwell and AV~Hill.
\newblock The velocity of the pulse wave in man.
\newblock {\em Proceedings of the Royal Society of London B: Biological
  Sciences}, 93:298--306, 1922.

\bibitem{kubicek1966}
W.~G. Kubicek, R.~P. Patterson, and D.~A. Witsoe.
\newblock Impedance cardiography as a noninvasive method of monitoring cardiac
  function and other parameters of the cardiovascular system.
\newblock {\em Annals of the New York Academy of Sciences}, 170(2):724--732,
  1966.

\bibitem{bernstein1986}
D.~P. Bernstein.
\newblock A new stroke volume equation for thoracic electrical bioimpedance:
  theory and rationale.
\newblock {\em Critical Care Medicine}, 14(10):904--909, 1986.

\bibitem{sugawara2010}
J~Sugawara, K~Hayashi, T~Yokoi, and H~Tanaka.
\newblock Carotid-femoral pulse wave velocity: impact of different arterial
  path length measurements.
\newblock {\em Artery Research}, 4:27--31, 2010.

\bibitem{hausvater2012}
A~Hausvater, T~Giannone, YH~Sandoval, et~al.
\newblock The association between preeclampsia and arterial stiffness.
\newblock {\em Journal of Hypertension}, 30:17--33, 2012.

\bibitem{ronnback2005}
M~R{\"o}nnback, K~Lampinen, PH~Groop, and R~Kaaja.
\newblock Pulse wave reflection in currently and previously preeclamptic women.
\newblock {\em Hypertension in Pregnancy}, 24:171--180, 2005.

\end{thebibliography}

\end{document}